# Terahertz Generation in Lithium Niobate Driven by Ti:Sapphire Laser Pulses and its Limitations

Xiaojun Wu,[1,*] Sergio Carbajo,[1,2] Koustuban Ravi,[3] Frederike Ahr,[1,2] Giovanni Cirmi,[1] Yue Zhou,[1] Oliver D. Mücke,[1] and Franz X. Kärtner[1,2,3]

[1]*Center for Free-Electron Laser Science, DESY and Center for Ultrafast Imaging, Hamburg 22607, Germany*
[2]*Department of Physics, University of Hamburg, Hamburg 22761, Germany*
[3]*Department of Electrical Engineering and Computer Science and Research Laboratory of Electronics, Massachusetts Institute of Technology, Cambridge, Massachusetts 02139, USA*
*Corresponding author: xiaojun.wu@desy.de*

We experimentally investigate the limits to 800 nm-to-terahertz (THz) energy conversion in lithium niobate at room temperature driven by amplified Ti:Sapphire laser pulses with tilted-pulse-front. The influence of the pump central wavelength, pulse duration, and fluence on THz generation is studied. We achieved a high peak efficiency of 0.12% using transform limited 150 fs pulses and observed saturation of the optical to THz conversion efficiency at a fluence of 15 mJ/cm$^2$. We experimentally identify two main limitations for the scaling of optical-to-THz conversion efficiencies: (i) the large spectral broadening of the optical pump spectrum in combination with large angular dispersion of the tilted-pulse-front and (ii) free-carrier absorption of THz radiation due to multi-photon absorption of the 800 nm radiation.

Strong-field terahertz (THz) sources hold promise for enabling myriads of novel applications. They possess intense electric and magnetic fields at frequencies which are particularly amenable to studies of condensed matter dynamics [1-3], manipulation of molecules [4], high-harmonic generation (HHG) [5], and compact charged-particle acceleration [6-7], among others. Therefore, there is a great need for the development of robust and efficient strong-field THz sources.

These sources are predominantly accelerator-based facilities (delivering up to 100 μJ THz energy) [8, 9] or ultrafast laser-based table-top systems [10]. Laboratory scale systems are of particular interest due to accessibility and relatively low cost. In this category, laser-induced air/gas plasmas (delivering up to 5 μJ THz energy) [11] and optical rectification (OR) of infrared (IR) pulses in nonlinear optical crystals (delivering up to 0.4 mJ THz energy) [12] have emerged as the most common methods of all THz generation modalities. The highest optical-to-THz conversion efficiencies (henceforth referred to as conversion efficiency) in excess of 1% at room temperature have been achieved by OR employing angularly dispersed femtosecond IR pump pulses in lithium niobate [13, 14]. As a result, these systems are especially relevant to generating mJ-level THz pulses [12]. In this approach, angular dispersion is introduced to compensate for the large difference in refractive indices between IR and THz frequencies by forming a tilted-pulse-front. The generated THz then propagates perpendicular to this tilted-pulse-front and phase matching is achieved in a non-collinear configuration. OR systems based on lithium niobate are powered predominantly by sources in the 1 μm and 800 nm wavelength regions. While the former yields much higher conversion efficiencies, today, 800 nm sources based on Ti:Sapphire systems prevail in ultrafast laser technology as the most accessible and widely employed sources. Consequently, exploring the limits to achievable conversion efficiency is a great value in the pursuit of accessible high-energy THz sources.

In this paper we extensively investigate the limits of conversion efficiency with 800 nm systems through a systematic study of all relevant pump pulse parameters: fluence, bandwidth, pulse-width, and pump central wavelength. We experimentally measured a maximum conversion efficiency of 0.12% with a pulse duration of 150 fs using congruent lithium niobate at room temperature. This trend is contrary to earlier theoretical predictions that the optimal conversion efficiency occurs at ~500 fs [15]. In addition, saturation of conversion efficiency at a fluence of 15 mJ/cm$^2$ was measured. The studies in Ref. [16-18] have suggested that free-carrier absorption (FCA) of THz radiation due to carriers generated by multi-photon absorption of 800 nm radiation as being an important limitation to scaling THz energies. In this work, we present experimental results depicting the effect of the pump central wavelength on conversion efficiency for the first time. We observe a linear 50% increase in conversion efficiency with a shift in pump central wavelength from 801 to 806 nm. This trend is correlated to that exhibited by the three-photon absorption (3PA) coefficient for lithium niobate in this wavelength range. Hence, these results are an indication of the detrimental effects posed by multi-photon absorption.

In Ref. [19], the saturation behavior was attributed to the phase mismatch caused by self-phase modulation (SPM) effects. In that work, SPM was recognized as the most important limitation to THz generation at large pump intensities. However, recent theoretical calculations [20] suggest that the large frequency down-shift and spectral broadening of the IR spectrum (cascading), which manifest as a consequence of efficient THz generation, in conjunction with the large angular dispersion in OR is a stronger limitation in comparison to SPM. In this paper, we experimentally verify this argument for 800 nm pumping by measuring the transmitted IR spectra. We observe that in the absence of THz generation, the broadening of the IR spectrum due to SPM effects is significantly smaller than in the presence of THz. Since the effects of phase mismatch are larger for larger bandwidths, this indicates that cascading effects are more important in comparison to SPM effects for the same pump intensities. Therefore, the limitations of THz generation using 800 nm Ti:Sapphire lasers are attributed to two main factors: cascading effects in conjunction with angular dispersion and FCA caused by multi-photon absorption of 800 nm pump radiation. The relative importance of the two effects remains to be fully explored and is a topic for future investigations.

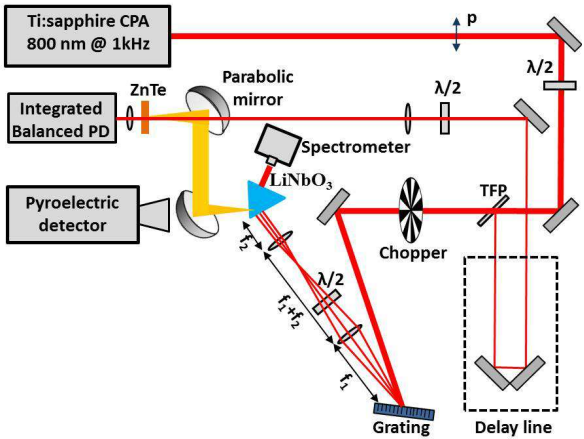

Fig. 1. (Color Online) Experimental setup for THz generation in lithium niobate via OR employing the pulse-front tilting technique. The focal lengths $f_1$ and $f_2$ are 150 mm and 50 mm, respectively (TFP: thin-film polarizer).

We constructed the THz generation system shown in Fig. 1. A Ti:Sapphire system (Coherent Legend Elite Duo and cryogenic single-pass power amplifier: 150 fs transform limited pulses at 800 nm wavelength and 1 kHz repetition rate) delivered up to 8 mJ pulses for this experiment. The output is divided into two beams by using a half-wave plate combined with a thin-film polarizer. The excitation beam passes through a mechanical chopper and is incident on a reflection grating with 1500 lines/mm at an incident angle of ~35°. The -1st order diffracted beam passes through a first lens ($f_1$=150 mm), a half-wave plate rotating the polarization to s, a second lens ($f_2$=50 mm), and is finally imaged into a congruent lithium niobate crystal. The lithium niobate crystal is doped with 6% MgO, and z-cut into an isosceles prism (57.9×57.9×54.4 mm$^3$ and 25.4 mm thick) with an apex angle of 56°. We measured the THz energy at the output face of the prism using a calibrated pyroelectric detector (Microtech Instruments) with a modulation frequency of 20 Hz. We detected the temporal waveform of the THz fields by electro-optic sampling in a ZnTe crystal. In order to systematically investigate the influence of both the pulse duration and the central wavelength on the conversion efficiency, a variable width slit was placed in the grating compressor of the Ti:Sapphire system, which allowed to tune the bandwidth and central wavelength of the pump pulses. We also monitored both, the spectral profiles of the input and transmitted pump laser pulses using an optical spectrum analyzer. We confirmed with an interferometric autocorrelator that the pulses used were transform limited (the lower limit for the pulse duration for a given pulse spectrum). Hence we calculated the pump pulse durations from the Fourier limited IR input spectra.

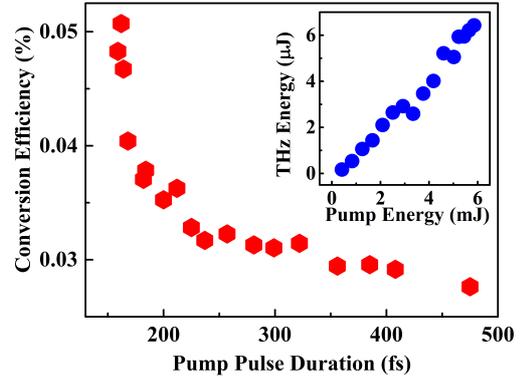

Fig. 2. (Color online) IR-to-THz conversion efficiency as a function of the pump pulse duration. $\lambda_c$=804 nm; pump fluence=6.5 mJ/cm$^2$. Inset: THz energy versus pump energy for $\tau$=157 fs.

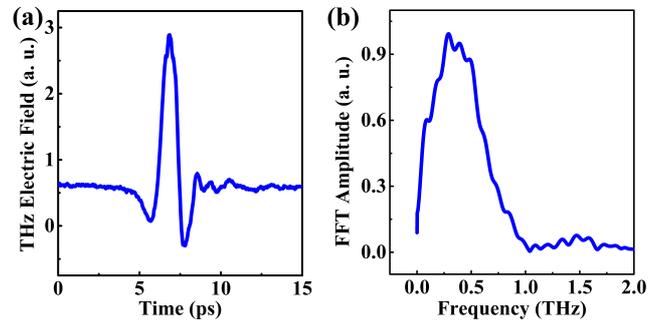

Fig. 3. (Color online) (a) THz temporal waveform measured by electro-optic sampling and (b) corresponding Fast Fourier Transform (FFT) amplitude.

In order to verify the optimal pulse duration for THz generation in lithium niobate, we measured the conversion efficiency dependence on the transform limited pump pulse duration using a constant pump fluence of 6.5 mJ/cm$^2$ and central wavelength of 804 nm, as shown in Fig. 2. The conversion efficiency decreased by 45% when the pump pulse duration was increased from 157 to 475 fs, which exhibits an opposite trend compared with the theoretical predictions in Ref. [15]. We achieved a maximum THz energy of 6.4 µJ (average power of 6.4 mW) for the shortest transform limited pulse duration of 157 fs and its corresponding maximum conversion efficiency of 0.12%, as depicted in the inset of Fig. 2. Fig. 3 shows the THz temporal waveform measured by electro-optic sampling from lithium niobate when excited by 157 fs pulses with 2 mJ pump energy and its corresponding spectrum. The THz pulses contain spectral components up to 1 THz and exhibit maximum spectral content at 0.35 THz.

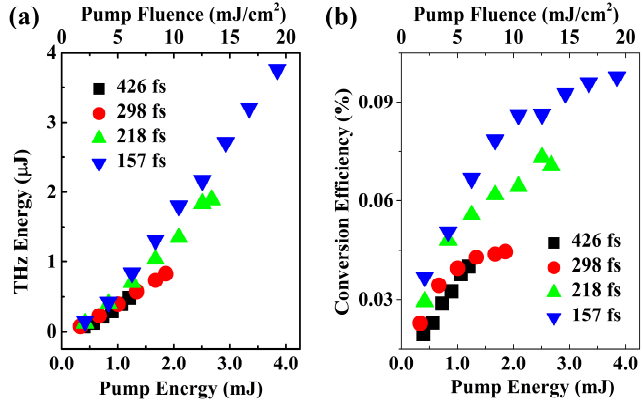

Fig. 4. (Color online) (a) Dependence of THz energy on pump energy (fluence) for pump pulse durations of τ=426, 298, 218, and 157 fs, respectively. (b) Corresponding IR-to-THz energy efficiency slope.

Fig. 4(a) shows the THz energy as a function of pump laser energy for different pulse width τ=426, 298, and 218 fs with $\lambda_c$=806 nm as well as for τ=157 fs with $\lambda_c$= 804 nm, respectively. The amount of generated THz energy monotonically increases with reduced pulse duration or, alternatively, increased peak intensity. Fig. 4(b) for τ=298, 218, and 157 fs show that the conversion efficiency dramatically increases first with pump energy and then shows slight saturation. The shortest pulse duration of 157 fs exhibits the highest saturation threshold of ~15.0 mJ/cm². Saturation is not observed for τ = 426 fs due to the limited pump energy from the variable width slit. In earlier studies, FCA due to multi-photon absorption was cited as the principal reason for THz energy saturation. However, in Ref. [20], the impact of FCA was negated and SPM was suggested to be the most important saturation mechanism. Here, we experimentally demonstrate two main limiting mechanisms for THz generation using 800 nm Ti:Sapphire systems. These include the combined action of the cascading effects with large angular dispersion, which is in agreement with recent theoretical calculations [20] and FCA absorption due to multi-photon absorption in lithium niobate.

Fig. 5 shows the emitted THz energy as function of the central wavelength ($\lambda_c$) of the pump pulses for a fixed pulse duration τ=298 fs and pump fluence of 7.6 mJ/cm². The generated THz energy monotonically increased with the central wavelength of the pump laser. When the central wavelength increased from 801 nm to 806 nm, the THz energy was enhanced by ~50%. This observation indicates that the central wavelength of the pump laser has a strong influence on the THz generation in this frequency range. We associate this dependence with 3PA. In fact, the direct band gap ($E_g$) of lithium niobate is in the range of 4 eV at room temperature and a significant nonlinear absorption rate may originate from 3-photon interband transitions, where the photon energy (ℏω) is ~1.55 eV, even at moderate intensities of up to a few tens of GW/cm². For a given radiation intensity, the interband 3PA scales approximately linearly with photon energy in the case 3ℏω > $E_g$ [21], which is the case for 3PA in lithium niobate with a radiation wavelength centered around 800 nm. This process limits the conversion efficiency by significantly increasing the free carrier density, which then strongly absorb THz radiation. In previous work [16], FCA was cited as an important factor for saturation of THz energy. However, in Ref. [19], it was suggested that FCA plays no significant role in the OR processes in lithium niobate and has no effect at all in THz energy saturation. Moreover, SPM was recognized to be the only factor in the saturation of THz energy. The results shown in Fig. 5 depict for the first time a strong correlation between pump photon energy distribution on generated THz radiation and is evidence of the role played by FCA arising from 3PA.

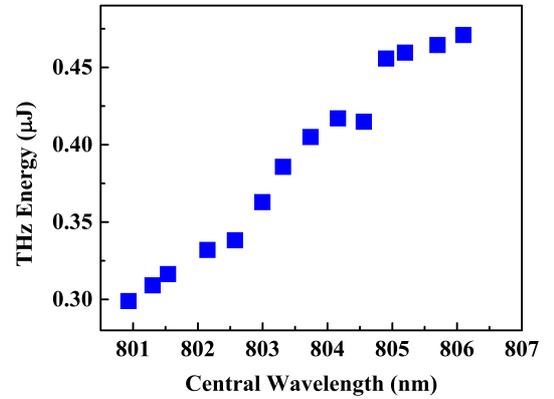

Fig. 5. (Color online) THz energy plotted as function of the pump central wavelength for Ti:Sapphire pump pulses with fixed pump intensity of 14 GW/cm², 5 nm bandwidth centered at 803 nm.

The key element to limiting the conversion efficiency in pulse-front-tilted OR is the short coherence length associated with strong angular dispersion in conjunction with cascaded broadening. In order to verify this constraint, we measured the spectral reshaping effects in the transmitted IR spectrum due to THz generation with and without phase matching by varying the IR input polarization. The measured transmitted IR spectra are shown in Fig. 6 for an input IR energy of ~5 mJ. For optimum polarization for THz generation, i.e. s-polarization, the spectrum was significantly red-shifted due to cascading, resulting in an output energy of 2 μJ. However, when the polarization was rotated to p-polarization, the resulting broadening was significantly less pronounced than that for s-polarization. Moreover, the THz output energy dropped a four-fold. Conversion efficiencies beyond the Manley-Rowe limit are intrinsically associated with cascading and thus it cannot be ignored when computing conversion efficiencies. From the recorded s-polarized broadened output spectrum, we estimate the number of photon cascading cycles ($N_{cas}$) to be ~14. Therefore, the intrinsic THz conversion efficiency is expected to be ~1.31%. The detriment in detected efficiency is primarily linked to FCA in the material, Fresnel losses, and finite detection. This experimental evidence verifies recent theoretical studies [20] which confirm that the inherently short coherence lengths of pulse-front tilted THz generation schemes due to angular dispersion are further shortened due to the phase-mismatch arising from the cascaded downshift of the pump pulse, and that these two are more important limiting agents in pulse-front tilted OR compared to SPM.

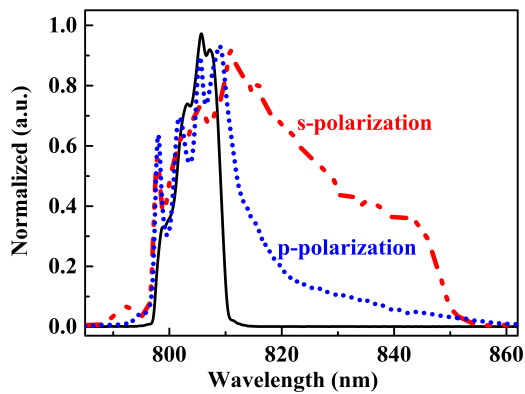

Fig. 6. (Color Online) Ti:Sapphire spectra before and after THz generation in lithium niobate crystal.

In conclusion, we have performed a comprehensive and parametric experimental investigation of the 800 nm-to-THz conversion efficiency in lithium niobate at room temperature. We experimentally measured a 0.12% optical-to-THz conversion efficiency using 150 fs transform limited pulses and observe a saturation fluence of 15 mJ/cm$^2$. We also show that, contrary to earlier studies, the cascading process together with strong angular dispersion and FCA of THz radiation due to 3PA of 800 nm radiation are the two main limiting mechanisms for pulse-front tilted OR schemes using Ti:Sapphire-based systems. Further improvement in conversion efficiencies can be achieved via cryogenic cooling and reduction of Fresnel losses.

The authors thank Guoqing (Noah) Chang, Ming Xin, Giulio M. Rossi, Shaobo Fang, Qing Zhang, and Shih-Hsuan Chia for their helpful discussion. This work has been supported by the excellence cluster 'The Hamburg Centre for Ultrafast Imaging - Structure, Dynamics and Control of Matter at the Atomic Scale' of the Deutsche Forschungsgemeinschaft and the Center for Free-Electron Laser Science at the Deutsches Elektronen-Synchrotron (DESY), a Center of the Helmholtz Association. Dr. Wu acknowledges support by Research Fellowship from the Alexander von Humboldt Foundation.